\documentclass[aps,prl,twocolumn,groupedaddress]{revtex4}

\usepackage{color}
\usepackage{graphicx}
\usepackage{amssymb, amsmath, amsfonts}

\begin{document}

\title{The light-well: A tuneable free-electron light source on a chip}

\author{G.~Adamo, K.~F.~MacDonald}
\email{kfm@orc.soton.ac.uk}
\homepage{www.nanophotonics.org.uk/niz}
\author{N.~I.~Zheludev}
\affiliation{Optoelectronics Research Centre, University of Southampton, Southampton, SO17 1BJ, UK}

\author{Y.~H.~Fu, C-M.~Wang and D.~P.~Tsai}
\affiliation{Department of Physics, National Taiwan University, Taipei 10617, Taiwan, Republic of China}

\author{F.~J.~Garc\'{\i}a~de~Abajo}
\affiliation{Instituto de \'{O}ptica - CSIC, Serrano 121, 28006 Madrid, Spain}

\date{\today}

\begin{abstract}
The passage of a free-electron beam through a nano-hole in a periodically layered metal/dielectric structure creates a new type of tuneable, nanoscale radiation source - a `light-well'. With a lateral size of just a few hundred nanometers, and an emission intensity of $\sim$200~W/cm$^2$ such light-wells may be employed in nanophotonic circuits as chip-scale sources, or in densely packed ensembles for optical memory and display applications.
\end{abstract}

\maketitle

With a view towards future highly-integrated nanophotonic devices, there is growing interest in nanoscale light and surface plasmon-polariton sources~\cite{Koller2008, Hill2007, Bergman2003, Zheludev2008a, Park2008}. Electron-beam-induced radiation emission~\cite{Egusa2004, Bashevoy2006, Wijngaarden2006} is of particular interest because electrons can be focused to nanoscale spots (far below the diffraction limit for light), enabling such applications as high-resolution mapping of plasmonic excitations in nanostructures~\cite{Bashevoy2007, Vesseur2007, Nelayah2007}. In this letter we report on the first demonstration of \emph{tuneable} light emission from a chip-scale free-electron-driven source - a `light-well' - in which optical (visible to near-infrared) photons are generated as an electron beam travels through a nano-hole in a layered metal-dielectric structure. With a lateral size of just a few hundred nanometres, and the notable advantage of wavelength tuneability, such structures may be employed in nanophotonic circuits as on-chip light sources, or in densely packed ensembles for optical memory and field emission or surface-conduction electron-emitter display applications~\cite{Nakamoto2008}.

The light-well belongs to a family of free-electron-driven emitters headed by the free-electron laser (FEL) \cite{O'Shea2001} - one of today's brightest and most widely tuneable radiation sources. In a FEL a beam of electrons from an accelerator passes through a magnetic undulator or `wiggler' that imposes periodic transverse oscillations on the electron trajectory, resulting in the coherent release of photons; The wavelength of emitted light can be tuned by adjusting the energy of the electron beam or the parameters of the undulator. The incoherent light source we report here may be seen as a nanoscale relative of the facility-scale FEL: In the light-well a beam of free electrons experiences an alternating dielectric environment as it travels through a tunnel in a periodically layered metal-dielectric nanostructure, and optical photons are emitted as a result; As in the FEL, the wavelength of radiation emitted by the light-well can be tuned by adjusting the energy of the electrons in the incident beam.

Experimental light-wells (see Fig.~1) were fabricated in a stack of eleven alternating silica and gold layers (six silica and five gold), each with a thickness of 200~nm, sputtered onto a thick gold base layer on a silicon substrate. Wells with typical diameters of around 700~nm were milled through the stack into the base layer, perpendicular to the plane of the layers, using a focused ion beam.

\begin{figure}
\includegraphics[width=80mm]{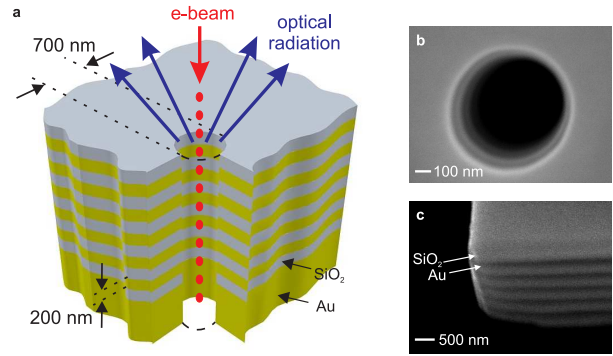}
\caption{(a) Schematic cut-away section of a light-well, which comprises a nano-hole through a stack of alternating metal and dielectric layers, into which an electron beam is launched. Light is generated as electrons travel down the well and encounter a periodic material environment. (b) Scanning electron microscope image of a light-well fabricated in a gold-silica multilayer. (c) The alternating metal-dielectric layer structure as seen at an exposed corner of the sample.}
\end{figure}

Characterization of the light-well as an optical source was performed in a scanning electron microscope (SEM) equipped with a light-collection system comprising a small parabolic mirror positioned above the sample (collecting emitted light over approximately half of the available hemispherical solid angle and directing it out of the SEM chamber) and a spectrograph equipped with a liquid-nitrogen-cooled CCD array detector. The electron beam of the SEM, focused to a spot with a diameter of $\sim$30~nm, was used to drive light-well optical emission. Emission spectra were recorded at different acceleration voltages and beam currents (measured using a Faraday cup in the SEM column) for a number of beam injection points along the diameter of the nano-hole, with reference measurements also taken for beams impacting the top (silica) surface of the multilayer outside the hole.

The following characteristic emission features have been observed:

1. When the electron beam hits the surface of the structure outside the hole, light emission is observed with a broad spectrum centered at 640~nm (Fig.~2a). Within experimental accuracy this spectrum does not depend on the electron acceleration voltage and is attributed simply to cathodoluminescence and backward transition radiation from the silica-gold multilayer structure~\cite{Yamamoto2001a, Koyama1980, Mooradian1969}.

2. When the electron beam is injected into the nano-hole the emission spectrum contains two peaks ($I$ and $II$)  with spectral positions that depend on the electron acceleration voltage (Fig.~2b). The emission intensity is found to increase as the injection point approaches the wall of the light-well as illustrated for peak $I$ in the inset to Fig.~2c.

3. For a fixed injection point and electron acceleration voltage, the emission intensity increases linearly with beam current (Fig.~2c).

\begin{figure}
\includegraphics[width=80mm]{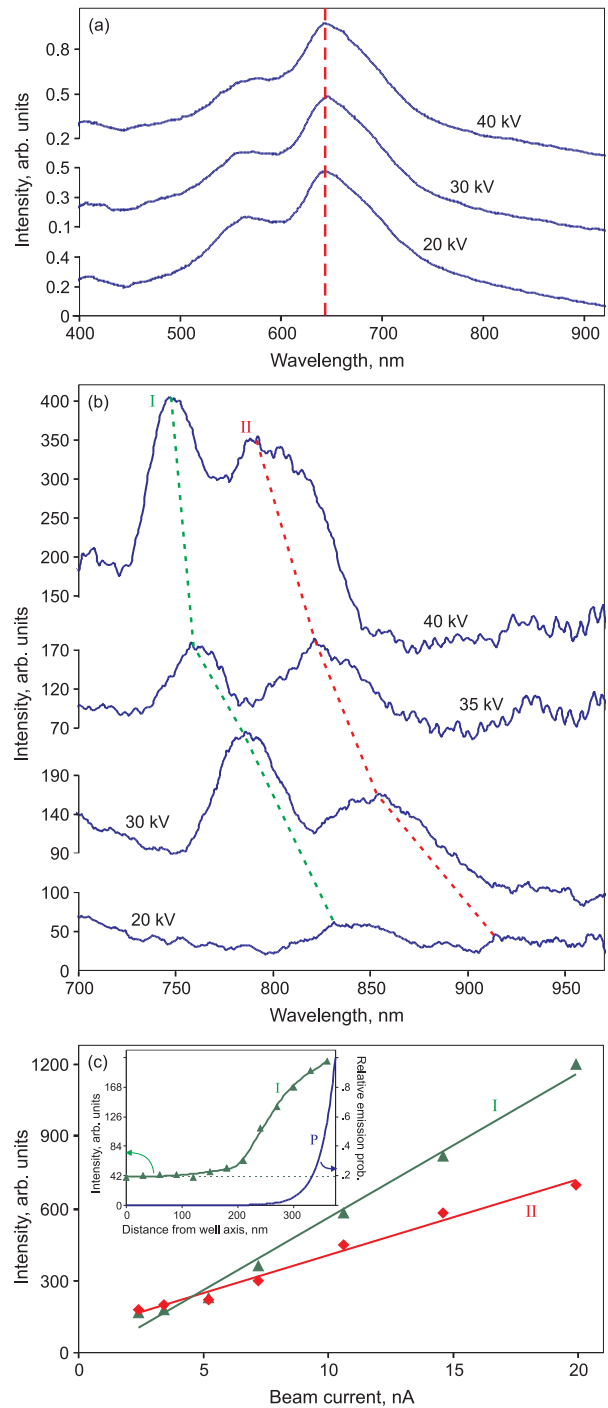}
\caption{(a) Emission spectra for an electron beam impact point outside the light-well, i.e. on the top surface of the gold-silica multilayer, for a range of electron acceleration voltages. (b) Emission spectra for a 750~nm diameter well for acceleration voltages of 20, 30, 35 and 40~kV with a beam injection point $\sim$100~nm inside the wall of the well. (c) Emission intensity for peaks $I$ and $II$ as a function of electron beam current for an acceleration voltage of 40 kV at an injection point $\sim$100~nm inside the wall of the well. The inset shows peak $I$ (760~nm) emission intensity at 40~kV and analytically estimated emission probability ($P$) as functions of beam injection position relative to the axis of a 750~nm diameter well.}
\end{figure}

Though structurally simple, the light-well's emission characteristics are controlled by a complex combination of material- and geometry-dependent processes that cannot presently be described within a single analytical or numerical model. To a first approximation, one may consider that light emission originates from an oscillating dipole source created as electrons experience a periodically modulated potential within the well. In this simplified picture, an electron passing through a metal section of the well interacts with its `mirror image', but this interaction is somewhat different in a dielectric section. This creates a monochromatic dipole of frequency $\nu = v/a$, where $v$ is the electron velocity and $a$ is the period of the structure, moving at the electron velocity. The radiation efficiency depends on the strength of the mirror interaction and on the density of photonic states available in the well, and therefore on its geometry and material composition as well as the proximity of electrons to the wall. In a well of finite length $L$, the spectral width of the emission line $\Delta\nu$ is governed by the uncertainty relation $L \times\Delta\nu \simeq v$. So for example, with 30~keV electrons ($v \simeq c/3$) and a well length $L = 2.2~\mu m$ (eleven 200~nm layers) one may expect an emission line with a width of $\sim$220~nm centered at $\sim$1200~nm, not far from what is observed in the experiment. In reality, this na\"{\i}ve picture is complicated by relativistic corrections, the excitation and scattering of surface plasmons on metal/dielectric interfaces within the structure, the light-guiding properties of the silica layers, and ultimately the guided-mode profile of the nano-tube (see below). Nevertheless, the inverse proportionality it describes between emission wavelength and electron velocity is clearly seen in the blue-shifting of experimentally observed emission peaks with increasing electron energy (Fig.~2b). For a well of ideal cylindrical symmetry the mirror interaction strength will be at a minimum for electrons traveling along the axis of the well and will increase with proximity to the wall, increasing the emission intensity as described above and illustrated (from experiment) in the inset to Fig.~2c. Also plotted here is an analytical estimate of the corresponding relative emission probability, which is proportional to $(I_m(\nu R / v \gamma))^2$, where $I_m$ is the modified Bessel function, $\gamma$ is the Lorentz correction factor $(1-v^2/c^2)^{-1}$, and $R$ is the radial distance of the beam injection point from the axis of the well. The offset zero level seen in the experimental data is related to the imperfect cylindrical symmetry of the well.

This basic emission process bears some similarity with the Smith-Purcell effect~\cite{Smith1953}, whereby a continuum of light wavelengths (varying with electron energy and the direction of emission) is generated as electrons pass over the surface of a planar metal grating. However, in the present case emission occurs via coupling to the 1D photonic bands of the periodically structured well so that discrete emission wavelengths are produced, determined by the condition that the wavevector of a guided mode is equal to $\nu/v$. This fundamental consideration is analyzed in Fig.~3, which shows the dispersion diagram for the guided electromagnetic modes of a cylindrical gold cavity with a radius of 350~nm, folded over the first Brillouin zone (BZ) to account for a periodicity $a$ of 400~nm in the direction parallel to the axis of the tube. Modes with azimuthal numbers $m$ = 0, 1 and 2 are shown alongside the free-space light line and lines associated with electron energies of 20, 30 and 40 keV. (For the same reason that periodic patterns of sub-wavelength holes in a planar metal surface do not substantially change its properties, even at high filing-factors, except very close to the BZ boundaries~\cite{Garc'iadeAbajo2007}, the presence of silica inclusions in the wall of the experimentally studied cavity will not significantly alter the mode structure except around $q$ = 0 and $\pi/a$ where some distortion will occur.) Electrons can couple to cavity modes where their respective lines intersect. This plot illustrates that the accessible mode profile is a complex function of electron energy, but one can immediately see that an electron of a given energy can couple to a number of modes at different energies (wavelengths) and that a given intersection point will shift to higher energies (shorter wavelengths) with increasing acceleration voltage, as observed experimentally.

\begin{figure}
\includegraphics[width=80mm]{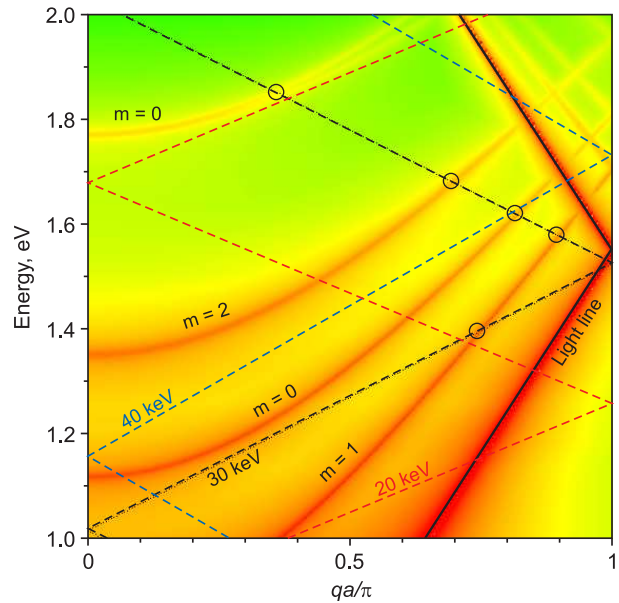}
\caption{Dispersion diagram showing the guided modes of an infinite periodic cylindrical gold cavity with a radius of 350 nm (azimuthal numbers $m$ = 0, 1 and 2), superimposed with the free-space light line (solid black) and lines associated with electron energies of 20, 30 and 40 keV (dashed lines). Points of intersection between the 30 keV line and the cavity modes are circled.}
\end{figure}

The efficiency of the emission process has been evaluated in terms of the number of photons generated per incident electron. (The number of photons being derived from a multiple gaussian fit to the recorded spectra, taking into account the throughput efficiency of the light collection system; the corresponding number of electrons being given by the beam current and sampling time.) It is found to reach $1.9\times10^{-5}$ for peak $I$ and $3.4\times10^{-5}$ for peak $II$ at 40~keV, giving a light source with an output power of the order of 0.1~nW in either case (an emission intensity of $\sim$200~W/cm$^2$). It is anticipated that optimization of the light-well geometry, material composition and pumping regime will substantially improve these figures, and that longer wells will show narrower emission lines. Indeed, if losses can be controlled (for example, by inhibiting surface plasmon generation and transverse light-guiding) and higher (perhaps pulsed) drive currents are employed, it may be possible to achieve superradiant or even lasing emission modes~\cite{Gover2005, Andrews2004, Bakhtyari2002}.

To summarize, we provide the first proof-of-concept demonstration of a tuneable, electron-beam-driven, nanoscale radiation source in which light is generated as free-electrons travel down a `light-well' - a nano-hole through a stack of alternating metal and dielectric layers. Near-infrared emission is demonstrated in the present case but the concept may readily be scaled to other wavelength ranges by varying the periodicity of the structure. The simplicity and nanoscale dimensions of the light-well geometry make it a potentially important device for future integrated nanophotonic circuit, optical memory and display applications where it may be driven by the kinds of microscopic electron sources already developed for ultrahigh-frequency nanoelectronics and next-generation flat-panel displays~\cite{Nakamoto2008}.

\begin{acknowledgments}
 The authors from the University of Southampton acknowledge the support of the UK Engineering and Physical Sciences Research Council (EPSRC, projects EP/C511786/1 and EP/F012810/1) and the European Union (FP6 project NMP4-2006-016881, `SPANS'); Those from the National Taiwan University thank the National Science Council of Taiwan (NSC-97-2120-M-002-013, NSC-97-2112-M-002-023-MY2, and NSC-97-2915-I-002-001) and the EPSRC (EP/F012810/1); Prof. Garc\'{\i}a~de~Abajo acknowledges the European Union (NMP4-2006-016881) and the Spanish MEC (MAT2007-66050 and Consolider `NanoLight.es').
\end{acknowledgments}


\begin{thebibliography}{21}
\expandafter\ifx\csname natexlab\endcsname\relax\def\natexlab#1{#1}\fi
\expandafter\ifx\csname bibnamefont\endcsname\relax
  \def\bibnamefont#1{#1}\fi
\expandafter\ifx\csname bibfnamefont\endcsname\relax
  \def\bibfnamefont#1{#1}\fi
\expandafter\ifx\csname citenamefont\endcsname\relax
  \def\citenamefont#1{#1}\fi
\expandafter\ifx\csname url\endcsname\relax
  \def\url#1{\texttt{#1}}\fi
\expandafter\ifx\csname urlprefix\endcsname\relax\def\urlprefix{URL }\fi
\providecommand{\bibinfo}[2]{#2}
\providecommand{\eprint}[2][]{\url{#2}}

\bibitem[{\citenamefont{Koller et~al.}(2008)\citenamefont{Koller, Hohenau,
  Ditlbacher, Galler, Reil, Aussenegg, Leitner, List, and Krenn}}]{Koller2008}
\bibinfo{author}{\bibfnamefont{D.~M.} \bibnamefont{Koller}},
  \bibinfo{author}{\bibfnamefont{A.}~\bibnamefont{Hohenau}},
  \bibinfo{author}{\bibfnamefont{H.}~\bibnamefont{Ditlbacher}},
  \bibinfo{author}{\bibfnamefont{N.}~\bibnamefont{Galler}},
  \bibinfo{author}{\bibfnamefont{F.}~\bibnamefont{Reil}},
  \bibinfo{author}{\bibfnamefont{F.~R.} \bibnamefont{Aussenegg}},
  \bibinfo{author}{\bibfnamefont{A.}~\bibnamefont{Leitner}},
  \bibinfo{author}{\bibfnamefont{E.~J.~W.} \bibnamefont{List}},
  \bibnamefont{and} \bibinfo{author}{\bibfnamefont{J.~R.} \bibnamefont{Krenn}},
  \bibinfo{journal}{Nat. Photon.} \textbf{\bibinfo{volume}{2}},
  \bibinfo{pages}{684} (\bibinfo{year}{2008}).

\bibitem[{\citenamefont{Hill et~al.}(2007)\citenamefont{Hill, Oei, Smalbrugge,
  Zhu, de~Vries, van Veldhoven, van Otten, Eijkemans, Turkiewicz, de~Waardt
  et~al.}}]{Hill2007}
\bibinfo{author}{\bibfnamefont{M.~T.} \bibnamefont{Hill}},
  \bibinfo{author}{\bibfnamefont{Y.-S.} \bibnamefont{Oei}},
  \bibinfo{author}{\bibfnamefont{B.}~\bibnamefont{Smalbrugge}},
  \bibinfo{author}{\bibfnamefont{Y.}~\bibnamefont{Zhu}},
  \bibinfo{author}{\bibfnamefont{T.}~\bibnamefont{de~Vries}},
  \bibinfo{author}{\bibfnamefont{P.~J.} \bibnamefont{van Veldhoven}},
  \bibinfo{author}{\bibfnamefont{F.~W.~M.} \bibnamefont{van Otten}},
  \bibinfo{author}{\bibfnamefont{T.~J.} \bibnamefont{Eijkemans}},
  \bibinfo{author}{\bibfnamefont{J.~P.} \bibnamefont{Turkiewicz}},
  \bibinfo{author}{\bibfnamefont{H.}~\bibnamefont{de~Waardt}},
  \bibnamefont{et~al.}, \bibinfo{journal}{Nat. Photon.}
  \textbf{\bibinfo{volume}{1}}, \bibinfo{pages}{589} (\bibinfo{year}{2007}).

\bibitem[{\citenamefont{Bergman and Stockman}(2003)}]{Bergman2003}
\bibinfo{author}{\bibfnamefont{D.~J.} \bibnamefont{Bergman}} \bibnamefont{and}
  \bibinfo{author}{\bibfnamefont{M.~I.} \bibnamefont{Stockman}},
  \bibinfo{journal}{Phys. Rev. Lett.} \textbf{\bibinfo{volume}{90}},
  \bibinfo{pages}{027402} (\bibinfo{year}{2003}).

\bibitem[{\citenamefont{Zheludev et~al.}(2008)\citenamefont{Zheludev,
  Prosvirnin, Papasimakis, and Fedotov}}]{Zheludev2008a}
\bibinfo{author}{\bibfnamefont{N.~I.} \bibnamefont{Zheludev}},
  \bibinfo{author}{\bibfnamefont{S.~L.} \bibnamefont{Prosvirnin}},
  \bibinfo{author}{\bibfnamefont{N.}~\bibnamefont{Papasimakis}},
  \bibnamefont{and} \bibinfo{author}{\bibfnamefont{V.~A.}
  \bibnamefont{Fedotov}}, \bibinfo{journal}{Nat. Photon.}
  \textbf{\bibinfo{volume}{2}}, \bibinfo{pages}{351} (\bibinfo{year}{2008}).

\bibitem[{\citenamefont{Park et~al.}(2008)\citenamefont{Park, Barrelet, Wu,
  Tian, Qian, and Lieber}}]{Park2008}
\bibinfo{author}{\bibfnamefont{H.-G.} \bibnamefont{Park}},
  \bibinfo{author}{\bibfnamefont{C.~J.} \bibnamefont{Barrelet}},
  \bibinfo{author}{\bibfnamefont{Y.}~\bibnamefont{Wu}},
  \bibinfo{author}{\bibfnamefont{B.}~\bibnamefont{Tian}},
  \bibinfo{author}{\bibfnamefont{F.}~\bibnamefont{Qian}}, \bibnamefont{and}
  \bibinfo{author}{\bibfnamefont{C.~M.} \bibnamefont{Lieber}},
  \bibinfo{journal}{Nat. Photon.} \textbf{\bibinfo{volume}{2}},
  \bibinfo{pages}{622} (\bibinfo{year}{2008}).

\bibitem[{\citenamefont{Egusa et~al.}(2004)\citenamefont{Egusa, Liau, and
  Scherer}}]{Egusa2004}
\bibinfo{author}{\bibfnamefont{S.}~\bibnamefont{Egusa}},
  \bibinfo{author}{\bibfnamefont{Y.-H.} \bibnamefont{Liau}}, \bibnamefont{and}
  \bibinfo{author}{\bibfnamefont{N.~F.} \bibnamefont{Scherer}},
  \bibinfo{journal}{Appl. Phys. Lett.} \textbf{\bibinfo{volume}{84}},
  \bibinfo{pages}{1257} (\bibinfo{year}{2004}).

\bibitem[{\citenamefont{Bashevoy et~al.}(2006)\citenamefont{Bashevoy, Jonsson,
  Krasavin, Zheludev, Chen, and Stockman}}]{Bashevoy2006}
\bibinfo{author}{\bibfnamefont{M.~V.} \bibnamefont{Bashevoy}},
  \bibinfo{author}{\bibfnamefont{F.}~\bibnamefont{Jonsson}},
  \bibinfo{author}{\bibfnamefont{A.~V.} \bibnamefont{Krasavin}},
  \bibinfo{author}{\bibfnamefont{N.~I.} \bibnamefont{Zheludev}},
  \bibinfo{author}{\bibfnamefont{Y.}~\bibnamefont{Chen}}, \bibnamefont{and}
  \bibinfo{author}{\bibfnamefont{M.~I.} \bibnamefont{Stockman}},
  \bibinfo{journal}{Nano Lett.} \textbf{\bibinfo{volume}{6}},
  \bibinfo{pages}{1113} (\bibinfo{year}{2006}).

\bibitem[{\citenamefont{van Wijngaarden et~al.}(2006)\citenamefont{van
  Wijngaarden, Verhagen, Polman, Ross, Lezec, and Atwater}}]{Wijngaarden2006}
\bibinfo{author}{\bibfnamefont{J.~T.} \bibnamefont{van Wijngaarden}},
  \bibinfo{author}{\bibfnamefont{E.}~\bibnamefont{Verhagen}},
  \bibinfo{author}{\bibfnamefont{A.}~\bibnamefont{Polman}},
  \bibinfo{author}{\bibfnamefont{C.~E.} \bibnamefont{Ross}},
  \bibinfo{author}{\bibfnamefont{H.~J.} \bibnamefont{Lezec}}, \bibnamefont{and}
  \bibinfo{author}{\bibfnamefont{H.~A.} \bibnamefont{Atwater}},
  \bibinfo{journal}{Appl. Phys. Lett.} \textbf{\bibinfo{volume}{88}},
  \bibinfo{pages}{221111} (\bibinfo{year}{2006}).

\bibitem[{\citenamefont{Bashevoy et~al.}(2007)\citenamefont{Bashevoy, Jonsson,
  MacDonald, and Zheludev}}]{Bashevoy2007}
\bibinfo{author}{\bibfnamefont{M.~V.} \bibnamefont{Bashevoy}},
  \bibinfo{author}{\bibfnamefont{F.}~\bibnamefont{Jonsson}},
  \bibinfo{author}{\bibfnamefont{K.~F.} \bibnamefont{MacDonald}},
  \bibnamefont{and} \bibinfo{author}{\bibfnamefont{N.~I.}
  \bibnamefont{Zheludev}}, \bibinfo{journal}{Opt. Exp.}
  \textbf{\bibinfo{volume}{15}}, \bibinfo{pages}{11313} (\bibinfo{year}{2007}).

\bibitem[{\citenamefont{Vesseur et~al.}(2007)\citenamefont{Vesseur, de~Waele,
  Kuttge, and Polman}}]{Vesseur2007}
\bibinfo{author}{\bibfnamefont{E.~J.~R.} \bibnamefont{Vesseur}},
  \bibinfo{author}{\bibfnamefont{R.}~\bibnamefont{de~Waele}},
  \bibinfo{author}{\bibfnamefont{M.}~\bibnamefont{Kuttge}}, \bibnamefont{and}
  \bibinfo{author}{\bibfnamefont{A.}~\bibnamefont{Polman}},
  \bibinfo{journal}{Nano Lett.} \textbf{\bibinfo{volume}{7}},
  \bibinfo{pages}{2843} (\bibinfo{year}{2007}).

\bibitem[{\citenamefont{Nelayah et~al.}(2007)\citenamefont{Nelayah, Kociak,
  St\'ephan, Garc\'ia~de Abajo, Tenc\'e, Henrard, Taverna, Pastoriza-Santos,
  Liz-Marz\'an, and Colliex}}]{Nelayah2007}
\bibinfo{author}{\bibfnamefont{J.}~\bibnamefont{Nelayah}},
  \bibinfo{author}{\bibfnamefont{M.}~\bibnamefont{Kociak}},
  \bibinfo{author}{\bibfnamefont{O.}~\bibnamefont{St\'ephan}},
  \bibinfo{author}{\bibfnamefont{F.~J.} \bibnamefont{Garc\'ia~de Abajo}},
  \bibinfo{author}{\bibfnamefont{M.}~\bibnamefont{Tenc\'e}},
  \bibinfo{author}{\bibfnamefont{L.}~\bibnamefont{Henrard}},
  \bibinfo{author}{\bibfnamefont{D.}~\bibnamefont{Taverna}},
  \bibinfo{author}{\bibfnamefont{I.}~\bibnamefont{Pastoriza-Santos}},
  \bibinfo{author}{\bibfnamefont{L.~M.} \bibnamefont{Liz-Marz\'an}},
  \bibnamefont{and} \bibinfo{author}{\bibfnamefont{C.}~\bibnamefont{Colliex}},
  \bibinfo{journal}{Nat. Phys.} \textbf{\bibinfo{volume}{3}},
  \bibinfo{pages}{348 } (\bibinfo{year}{2007}).

\bibitem[{\citenamefont{Nakamoto}(2008)}]{Nakamoto2008}
\bibinfo{author}{\bibfnamefont{M.}~\bibnamefont{Nakamoto}}, in
  \emph{\bibinfo{booktitle}{IEEE Industry Applications Society Annual Meeting}}
  (\bibinfo{publisher}{IEEE}, \bibinfo{address}{Edmonton, Alberta, Canada},
  \bibinfo{year}{2008}), pp. \bibinfo{pages}{447--451}.

\bibitem[{\citenamefont{O'Shea and Freund}(2001)}]{O'Shea2001}
\bibinfo{author}{\bibfnamefont{P.~G.} \bibnamefont{O'Shea}} \bibnamefont{and}
  \bibinfo{author}{\bibfnamefont{H.~P.} \bibnamefont{Freund}},
  \bibinfo{journal}{Science} \textbf{\bibinfo{volume}{292}},
  \bibinfo{pages}{1853} (\bibinfo{year}{2001}).

\bibitem[{\citenamefont{Yamamoto et~al.}(2001)\citenamefont{Yamamoto, Araya,
  Toda, and Sugiyama}}]{Yamamoto2001a}
\bibinfo{author}{\bibfnamefont{N.}~\bibnamefont{Yamamoto}},
  \bibinfo{author}{\bibfnamefont{K.}~\bibnamefont{Araya}},
  \bibinfo{author}{\bibfnamefont{A.}~\bibnamefont{Toda}}, \bibnamefont{and}
  \bibinfo{author}{\bibfnamefont{H.}~\bibnamefont{Sugiyama}},
  \bibinfo{journal}{Surf. Interface Anal.} \textbf{\bibinfo{volume}{31}},
  \bibinfo{pages}{79 } (\bibinfo{year}{2001}).

\bibitem[{\citenamefont{Koyama}(1980)}]{Koyama1980}
\bibinfo{author}{\bibfnamefont{H.}~\bibnamefont{Koyama}}, \bibinfo{journal}{J.
  Appl. Phys.} \textbf{\bibinfo{volume}{51}}, \bibinfo{pages}{2228}
  (\bibinfo{year}{1980}).

\bibitem[{\citenamefont{Mooradian}(1969)}]{Mooradian1969}
\bibinfo{author}{\bibfnamefont{A.}~\bibnamefont{Mooradian}},
  \bibinfo{journal}{Phys. Rev. Lett.} \textbf{\bibinfo{volume}{22}},
  \bibinfo{pages}{185} (\bibinfo{year}{1969}).

\bibitem[{\citenamefont{Smith and Purcell}(1953)}]{Smith1953}
\bibinfo{author}{\bibfnamefont{S.~J.} \bibnamefont{Smith}} \bibnamefont{and}
  \bibinfo{author}{\bibfnamefont{E.~M.} \bibnamefont{Purcell}},
  \bibinfo{journal}{Phys. Rev.} \textbf{\bibinfo{volume}{92}},
  \bibinfo{pages}{1069} (\bibinfo{year}{1953}).

\bibitem[{\citenamefont{Garc\'ia~de Abajo}(2007)}]{Garc'iadeAbajo2007}
\bibinfo{author}{\bibfnamefont{F.~J.} \bibnamefont{Garc\'ia~de Abajo}},
  \bibinfo{journal}{Rev. Mod. Phys.} \textbf{\bibinfo{volume}{79}},
  \bibinfo{pages}{1267} (\bibinfo{year}{2007}).

\bibitem[{\citenamefont{Gover}(2005)}]{Gover2005}
\bibinfo{author}{\bibfnamefont{A.}~\bibnamefont{Gover}},
  \bibinfo{journal}{Phys. Rev. ST Accel. Beams} \textbf{\bibinfo{volume}{8}},
  \bibinfo{pages}{030701} (\bibinfo{year}{2005}).

\bibitem[{\citenamefont{Andrews and Brau}(2004)}]{Andrews2004}
\bibinfo{author}{\bibfnamefont{H.~L.} \bibnamefont{Andrews}} \bibnamefont{and}
  \bibinfo{author}{\bibfnamefont{C.~A.} \bibnamefont{Brau}},
  \bibinfo{journal}{Phys. Rev. ST Accel. Beams} \textbf{\bibinfo{volume}{7}},
  \bibinfo{pages}{070701} (\bibinfo{year}{2004}).

\bibitem[{\citenamefont{Bakhtyari et~al.}(2002)\citenamefont{Bakhtyari, Walsh,
  and Brownell}}]{Bakhtyari2002}
\bibinfo{author}{\bibfnamefont{A.}~\bibnamefont{Bakhtyari}},
  \bibinfo{author}{\bibfnamefont{J.~E.} \bibnamefont{Walsh}}, \bibnamefont{and}
  \bibinfo{author}{\bibfnamefont{J.~H.} \bibnamefont{Brownell}},
  \bibinfo{journal}{Phys. Rev. E} \textbf{\bibinfo{volume}{65}},
  \bibinfo{pages}{066503} (\bibinfo{year}{2002}).

\end{thebibliography}

\end{document}